\documentclass[a4paper,12pt]{article}

\usepackage[english]{babel}
\usepackage{amsmath}
\usepackage{amsfonts}
\usepackage{amssymb}
\usepackage{a4wide}
\usepackage{graphicx}
\usepackage{cite}

\renewcommand{\not}[1]{#1 \hskip-0.475em /}

\begin{document}
\allowdisplaybreaks
\thispagestyle{empty}

\begin{flushright}
{\small
TUM-HEP-870/12\\
TTK-12-49\\
December 12, 2012}
\end{flushright}

\vskip1.5cm
\begin{center}
{\Large\bf\boldmath 
On ``dynamical mass'' generation in 
Euclidean de~Sitter space}
\vspace{\baselineskip}

\vspace{1cm}
{\sc M.~Beneke}${}^{a,b}$ 
and 
{\sc P.~Moch}${}^{a,b}$\\[5mm]
${}^a${\it Physik Department T31,\\ 
Technische Universit\"at M\"unchen,\\ 
James-Franck-Stra\ss e~1, D - 85748 Garching, Germany\\[0.3cm]
${}^b$
{\it Institut f{\"u}r Theoretische Teilchenphysik 
 und Kosmologie,\\ RWTH Aachen University,}\\
{\it D--52056 Aachen, Germany}
}\\[0.5cm]

\vspace*{1cm}
\textbf{Abstract}\\
\vspace{1\baselineskip}
\parbox{0.9\textwidth}{
We consider the perturbative treatment of the minimally 
coupled, massless, self-interacting scalar field in Euclidean de 
Sitter space. Generalizing work of Rajaraman, 
we obtain the dynamical mass $m^2\propto \sqrt{\lambda} H^2$ of 
the scalar for non-vanishing Lagrangian masses and the 
first perturbative quantum correction in the massless case. We develop the 
rules of a systematic perturbative expansion, which treats the 
zero-mode non-perturbatively, and goes in powers of $\sqrt{\lambda}$. 
The infrared divergences are self-regulated by the zero-mode dynamics.
Thus, in Euclidean de Sitter space the interacting, massless scalar field 
is just as well-defined as the massive field. 
We then show that the dynamical mass can be recovered from the diagrammatic 
expansion of the self-energy and a consistent solution of the 
Schwinger-Dyson equation, but requires the summation of a divergent  
series of loop diagrams of arbitrarily high order. Finally, 
we note that the value of the long-wavelength mode two-point 
function in Euclidean de Sitter space agrees at leading order 
with the stochastic treatment in Lorentzian de Sitter space, 
in any number of dimensions.}
\end{center}

\newpage
\setcounter{page}{1}


\newpage

\subsection*{1 Introduction}

It is well-known that the free, massless, minimally coupled scalar 
field in the de~Sitter background space-time cannot be defined in 
a de~Sitter invariant way due to infrared (IR) 
divergences~\cite{Allen:1985ux}. This occurs in any number of 
dimensions $d$, since the mode integral $\int d^{d-1}\vec{k} 
\,|\phi_k|^2$ that defines the Wightman two-point function 
always diverges logarithmically for $|\vec{k}\,|\to 0$.
In the expanding cosmological coordinate frame the divergence 
arises from the red-shifting of modes, which leads to a pile-up 
of long-distance modes at late times. But non-interacting 
fields are not very interesting. The question arises whether 
upon turning on an interaction of the scalar quantum field, no matter 
how small, the IR problem could somehow cure itself. For this 
to happen, the interaction must effectively become non-perturbatively 
strong among the long-distance modes. If so, the non-perturbative 
dynamics may be too complicated to be solved with analytic 
methods. However, it may also be that after a suitable 
resummation or reorganization of the expansion in the interaction 
strength, the interacting, massless, minimally coupled 
scalar field lends itself to a well-defined, systematic 
treatment.

Various previous results suggest that in scalar field theory with 
a quartic self-interaction $-\lambda\phi^4/4!$ the originally 
massless field acquires a dynamical mass $m^2_{\rm dyn} \propto 
\sqrt{\lambda} H^2$, where $H$ is the Hubble constant of 
de~Sitter space-time, which indeed regularizes the IR divergence.  
Starobinsky and Yokoyama~\cite{Starobinsky:1994bd} treat the 
long-distance fluctuations of the field as a classical 
random field that satisfies a Langevin equation. The associated 
Fokker-Planck equation is solved for large times by a 
probability distribution that results in finite correlation 
functions. Another approach uses the Schwinger-Dyson equations 
and obtains the dynamical mass from a self-consistent 
solution. In the mean-field or the large-$N$ 
limit~\cite{Riotto:2008mv,Serreau:2011fu} the self-energy 
can be restricted to the one-loop, tadpole diagram.
Garbrecht and Rigopoulos~\cite{Garbrecht:2011gu} analyzed the 
various in-in propagators in the CTP formalism and found 
that the large-$N$ result is modified by the two-loop self-energy, 
but remarkably, no further contribution arises beyond 
two loops due to systematic cancellations in the CTP 
index sums. However, while both formalisms agree on the parametric 
size $\sqrt{\lambda} H^2$ of the dynamical mass squared, the 
two exact results from \cite{Starobinsky:1994bd,Garbrecht:2011gu} 
disagree on the numerical prefactor. Neither of the formalisms 
so far explains how to compute sub-leading terms 
systematically.

The present work is motivated by the attempt to resolve the 
difference between the classical stochastic and 
diagrammatic/Schwinger-Dyson  
approach. For reasons that will become evident it is much simpler 
but still instructive to investigate the issue in Euclidean 
de~Sitter space, which is simply the sphere $S^d$. In an 
elegant paper Rajaraman~\cite{Rajaraman:2010xd} considered the 
functional integral on the sphere and identified the zero-mode 
integral as the origin of non-perturbative dynamics. He computed 
the two-point function of the zero mode, which can be related to
the dynamical mass. In this paper we extend the functional-integral 
approach and use 2PI methods to compute the exact self-energy, 
which is the central quantity in the diagrammatic approach. 
Our main results are as follows: 
\begin{itemize}
\item We formulate the rules for a well-defined perturbation 
expansion of correlation functions of the massless, minimally coupled 
scalar field in Euclidean de~Sitter space. The expansion parameter 
turns out to be $\sqrt{\lambda}$ instead of the coupling 
$\lambda$ of the standard perturbation expansion.
\item For the massive scalar field we obtain the dependence 
of the dynamical mass on the Lagrangian mass $m$; for the 
massless field the leading ${\cal O}(\sqrt{\lambda})$ correction.
\item We show that the dynamical mass can be obtained from the 
loop expansion of the self-energy after summing a divergent 
series to all orders in the loop expansion.
\item It seems to have gone unnoticed that the Euclidean 
dynamical mass~\cite{Rajaraman:2010xd} agrees with 
the result from the stochastic approach. We show that this is 
true in an arbitrary number of space-time dimensions despite 
the fact that the relevant dimension-dependent quantities 
are apparently unrelated.
\end{itemize} 
The interacting, massless, minimally coupled scalar field is 
therefore perfectly well-defined on the de~Sitter background. 
For $\lambda\ll 1$ there is a systematic weak-coupling 
expansion. The reason why this is possible despite the fact 
that the zero mode is truly strongly coupled is that the 
infrared theory consists of a single degree of freedom (the zero mode), 
whose dynamics can be solved exactly. What all this implies for Lorentzian 
de~Sitter space is less clear. We must leave this important point 
to further investigation.

\subsection*{2 Scalar field in 
Euclidean de~Sitter space}

Euclidean de~Sitter space is obtained from $d$-dimensional de~Sitter 
space in global coordinates with line element
\begin{equation}
ds^2= dt^2 - \frac{1}{H^2} \,\cosh^2(Ht)\,d\Omega_{d-1}
\end{equation}
by defining $t=-\frac{i}{H}\left(\theta-\frac{\pi}{2}\right)$  
and assuming $2\pi i/H$ periodicity in $t$, which turns 
$ds^2$ into 
\begin{equation}
ds^2 = -g_{\mu\nu} dx^\mu dx^\nu = 
-\frac{1}{H^2} \left(d\theta^2 + \sin^2\theta\,d\Omega_{d-1}\right).
\label{dsphere}
\end{equation}
Thus, Euclidean de~Sitter space is equivalent to the $d$-dimensional 
sphere with radius $1/H$. Because the sphere is compact, functions 
admit a discrete mode expansion in spherical harmonics. In $d$ 
dimensions the spherical harmonics $Y_{\vec L\,}(x)$ are labelled 
by the integer index vector $\vec L=(L,L_{d-1}, \ldots, L_1)$ with 
$L\geq L_{d-1} \geq \ldots \geq L_2 \geq |L_1|$ and satisfy 
(see, e.g., \cite{Higuchi:1986wu})
\begin{equation}
\Box \,Y_{\vec L} = \frac{1}{\sqrt{g}}\,\partial_\nu 
(\sqrt{g} g^{\nu\mu}\partial_\mu\,Y_{\vec L}) = - H^2 L (L+d-1) Y_{\vec L}
\end{equation}
as well as the orthogonality relation
\begin{equation}
\int d^d x\sqrt{g(x)}\, Y^*_{\vec L}(x) Y_{{\vec L}^\prime}(x)
 = \frac{1}{H^d}\,\delta_{{\vec L}{\vec L}^\prime}.
\label{orthogonality}
\end{equation}
The volume of Euclidean de~Sitter space is
\begin{equation}
V_d = \int d^d x\sqrt{g(x)} = \frac{2\pi^{\frac{d+1}{2}}}
{\Gamma(\frac{d+1}{2}) H^d} = \frac{1}{Y_0^2 H^d}.
\label{volume}
\end{equation}
Here $Y_0$ denotes the lowest harmonic, which is constant.

We consider the minimally coupled, real scalar field with 
Euclidean action
\begin{eqnarray}
S &=& \int d^d x \sqrt{g} \left\{\frac{1}{2}\,g^{\mu\nu} 
\partial_\mu \phi\partial_\nu\phi +\frac{1}{2} m^2 \phi^2 
+\frac{\lambda}{4!}\,\phi^4
\right\}
\nonumber\\
&=& \frac{1}{2} \sum_{\vec{L}} \frac{1}{H^d}\,
(H^2 L(L+d-1)+m^2)|\tilde\phi_{\vec L}|^2 +S_{\rm int}, 
\label{action}
\end{eqnarray}
where the second line follows from the mode expansion
\begin{equation}
\phi(x) = \sum_{\vec{L}} \tilde\phi_{\vec L}\,Y_{\vec L}(x)
\end{equation}
and the orthogonality relation (\ref{orthogonality}). From the quadratic 
terms of (\ref{action}) we deduce the free propagator
\begin{eqnarray}
G_{\rm free}(x,x^\prime) &=&
\sum_{\vec{L},\vec{L}^\prime} Y_{\vec L}(x) \,G_{\vec{L} \vec{L}^\prime}
Y^*_{\vec{L}^\prime}(x^\prime) 
= \sum_{\vec{L}}H^d\,\frac{Y_{\vec L}(x) Y^*_{\vec{L}}(x^\prime)}
{H^2 L(L+d-1)+m^2}
\nonumber\\
&=& \frac{H^{d-2}}{(4\pi)^{d/2}}\,
\frac{\Gamma(\frac{d-1}{2}+\nu)\Gamma(\frac{d-1}{2}-\nu)}
{\Gamma(\frac{d}{2})}\,
{}_2F_1\left(\frac{d-1}{2}+\nu,\frac{d-1}{2}-\nu;\frac{d}{2};
1-\frac{y}{4}\right)
\qquad
\label{propagator}
\end{eqnarray}
with
\begin{equation}
y = 2\left(1-\cos\theta\cos\theta^\prime-\sin\theta\sin\theta^\prime \, 
\vec{w}\cdot\vec{w}^\prime\,\right)
\quad \mbox{and} \quad
\nu = \sqrt{\left(\frac{d-1}{2}\right)^2-\frac{m^2}{H^2}}\,.
\end{equation}
Here $y$ is the invariant distance on the $d$-sphere $S^d$, and 
$\vec{w}$, $\vec{w}^\prime$ are two unit vectors on the sub-sphere 
$S^{d-1}$ with solid angle element $d\Omega_{d-1}$ in (\ref{dsphere}).
The second line of (\ref{propagator}) is indeed the de~Sitter 
propagator in the Bunch-Davies vacuum \cite{Candelas:1975du,Marolf:2010zp} in 
imaginary time.

The free propagator is ill-defined for $m=0$. The leading term 
for small $m$ is
\begin{equation}
G_{\rm free}(x,x^\prime) 
\stackrel{m\to 0}{\rightarrow} \frac{H^{d}}{(4\pi)^{d/2}}\,
\frac{\Gamma(d)}{\Gamma(\frac{d}{2})}\times \frac{1}{m^2}, 
\label{smallmasymptotics}
\end{equation}
which, as can be seen from the first line of (\ref{propagator}), 
originates only from the zero mode. Let us separate the constant 
zero mode from the field by defining
\begin{equation}
\phi(x) = \phi_0 + \hat\phi(x).
\end{equation}
The free propagator is the sum of the zero mode and non-zero mode 
propagator, since cross terms vanish by angular momentum conservation. 
The free zero-mode propagator 
equals the right-hand side of (\ref{smallmasymptotics}) for any 
$m=0\hspace*{-0.5cm}\not$\hspace*{0.4cm}, 
while the non-zero-mode propagator has a well-defined massless limit. 
In $d=4$, 
\begin{equation}
\hat{G}_{\rm free}(x,x^\prime) = \frac{H^2}{4\pi^2}
\left(\frac{1}{y}-\frac{1}{2}\ln\frac{y}{4}-1\right) \qquad (m=0).
\label{nonzeroprop}
\end{equation}

From now on we consider the massless, scalar field, $m=0$, unless 
mentioned otherwise, and assume $\lambda\ll 1$.  
The free zero-mode propagator is not defined, which is not 
surprising, since the zero mode has no quadratic term in the 
action (\ref{action}). The zero-mode must be treated non-perturbatively. 
Let 
\begin{equation} 
G(x,x^\prime) = G_0 + \hat G(x,x^\prime) = 
\langle \phi_0\phi_0\rangle+\langle \hat\phi(x)\hat\phi(x^\prime)\rangle
\end{equation}
be the exact two-point function of the interacting theory. Since 
$G_0$ is constant, we may write 
\begin{equation}
G_0 \equiv  \frac{H^d Y_0^2}{m_{\rm dyn}^2} = 
\frac{\Gamma(\frac{d+1}{2}) H^d}{2\pi^{\frac{d+1}{2}} m_{\rm dyn}^2}
= \frac{1}{V_d m_{\rm dyn}^2}.
\label{mdyndef}
\end{equation}
Comparison with the $L=0$ term in  the first line of (\ref{propagator}) 
suggests that we identify $m_{\rm dyn}^2$ with the dynamical mass of 
the originally massless scalar field, generated by the 
self-interaction. Note that this interpretation should be regarded 
with some caution, since the value of $m_{\rm dyn}^2$ is not 
related to the decrease of correlation functions at large separation 
$1/m_{\rm dyn}$. In fact, $1/m_{\rm dyn}$ corresponds to distances 
parametrically larger than the radius of the sphere, which carry no 
meaning. Similarly, in Lorentzian de Sitter space a dynamical mass 
of order $\lambda^{1/4} H\ll H$ is related to super-horizon correlations. 
Nevertheless, as will be seen below, 
a finite value of $G_0$ regularizes the IR divergence of the 
massless field and allows us to define a well-behaved perturbation 
expansion.

\subsection*{3 Perturbation expansion on the sphere}

In~\cite{Rajaraman:2010xd} the zero-mode two-point function  
$\langle \phi_0\phi_0\rangle$ was computed by evaluating the dominant 
contribution to its functional-integral representation, which 
gives
\begin{equation}
m_{\rm dyn}^2 = \frac{\Gamma(\frac{1}{4})}{\sqrt{4!} \,\Gamma(\frac{3}{4})}
\,\sqrt{\frac{1}{H^d V_d}}\,\sqrt{\lambda} H^{d/2}.
\label{mdyn}
\end{equation} 
In the following we generalize this approach. We show that both, 
$\langle \phi_0\phi_0\rangle$ and 
$\langle \hat\phi(x)\hat\phi(x^\prime)\rangle$ have well-defined 
perturbation expansions in $\sqrt{\lambda}$, and provide a set of 
Feynman rules for this expansion.

The generating functional is conveniently written in terms of 
two separate sources $J_0$, $\hat{J}(x)$, for the zero- and 
non-zero-mode field, respectively:
\begin{eqnarray}
Z[J_0,\hat{J}] &=& N \int {\cal D}[\phi_0]{\cal D}[\hat \phi] \,
\exp\left(-S - \int d^d x\sqrt{g}\,(J_0 \phi_0 +\hat{J}\hat{\phi})\right)
\nonumber\\
&=& \exp\left(-S_{\rm int}\!\left[\frac{\delta}{\delta J_0},
\frac{\delta}{\delta \hat{J}}\right]\right)\,
Z_0[J_0] \,\hat{Z}_{\rm free}[\hat{J}],
\label{Zpert}
\end{eqnarray}
where $N$ is defined such that $Z[0,0]=1$. Here 
\begin{equation}
S_{\rm int}[\phi_0,\hat{\phi}] = \frac{\lambda}{4!} \int d^dx\sqrt{g}
\left(\hat{\phi}^4 + 
4 \phi_0\hat{\phi}^3 + 
6 \phi_0^2\,\hat{\phi}^2 \right),
\end{equation}
and the term proportional to $\phi_0^3\hat\phi$ vanishes since 
$\int d^d x\sqrt{g}\, Y_{\vec{L}}(x) =0$ for $L>0$. 
The key point is that the term $\frac{\lambda}{4!}\,\phi_0^4$ is not 
included in $S_{\rm int}$, but must be part of $Z_0[J_0]$, 
since in the absence of a mass term for the scalar field the quadratic 
term in the zero-mode action vanishes, and 
the integral over $\phi_0$ in $Z_0[J_0]$ does not 
converge for large field values~\cite{Rajaraman:2010xd}. Hence,
\begin{equation}
Z_0[J_0] = N_0 \int {\cal D}[\phi_0]\,
\exp\left(- \int d^d x\sqrt{g}
\left(\frac{\lambda}{4!}\,\phi_0^4 +J_0 \phi_0\right)
\right),
\label{Z0}
\end{equation}
while the generating functional for the free non-zero-mode field,  
\begin{eqnarray}
\hat{Z}_{\rm free}[\hat{J}] &=& \hat{N}  \int {\cal D}[\hat \phi]\, 
\exp\left(- \int d^d x\sqrt{g}  
\,\left(\frac{1}{2}\,g^{\mu\nu} 
\partial_\mu \hat{\phi}\partial_\nu\hat{\phi} + \hat J \hat\phi
\right)\right)
\nonumber\\
&=& \exp\left(\frac{1}{2} \int d^d x\sqrt{g} \int d^d y\sqrt{g} \,
\hat{J}(x) \hat{G}_{\rm free}(x,y) \hat{J}(y)\right),
\end{eqnarray}
is a standard Gaussian functional integral. The zero-mode functional 
integral is simply an ordinary one-dimensional integral. Moreover, 
$\phi_0$ and $J_0$ are independent of $x$, so $\int d^d x\sqrt{g}=V_d$ 
in (\ref{Z0}).  The integral over $\phi_0$ can be evaluated exactly. 
Introducing 
\begin{equation}
\tilde\lambda = \frac{\lambda}{4!} V_d, \qquad 
\tilde{J}_0 = J_0 V_d,
\label{deftilde}
\end{equation}
we find
\begin{eqnarray}
Z_0[J_0] &=&  N_0 \int_{-\infty}^\infty d\phi_0\,
\exp\left(- \tilde{\lambda}\,\phi_0^4 - \tilde{J}_0 \phi_0\right)
\nonumber\\
&=&{}_0F_2\!\left(\frac{1}{2},\frac{3}{4};
\frac{\tilde{J}_0^4}{256 \tilde{\lambda}}\right) + 
\frac{\Gamma\left(\frac{3}{4}\right)}{2\Gamma\left(\frac{1}{4}\right)}
\frac{\tilde{J}_0^2}{\sqrt{\tilde{\lambda}}}\, \,
{}_0F_2\!\left(\frac{5}{4},\frac{3}{2};
\frac{\tilde{J}_0^4}{256 \tilde{\lambda}}\right),
\label{Z0integrated}
\end{eqnarray}
where ${}_0F_2$ denotes a hypergeometric function. One easily checks that 
\begin{equation}
\langle \phi_0^2\rangle_0 = \frac{\delta^2}{\delta \tilde{J}_0^2}\,
Z_0[J_0]_{|J_0=0} = \frac{1}{\sqrt{\tilde{\lambda}}}\,
\frac{\Gamma\left(\frac{3}{4}\right)}{\Gamma\left(\frac{1}{4}\right)}
\label{phisquared}
\end{equation}
reproduces (\ref{mdyn}) as is should be. Similarly, $\langle \phi_0^{2n}
\rangle_0$ follows from taking the appropriate number of derivatives. 
The index 0 on the bracket means that the computation is done with 
the zero-mode functional $Z_0$ alone. The full zero-mode $n$-point 
functions computed from $Z$ in (\ref{Zpert}) receive sub-leading 
corrections, as discussed below.

It is now straightforward to develop a systematic perturbative 
expansion of (\ref{Zpert}) and the corresponding Feynman rules. 
From (\ref{Z0}) it follows that every zero-mode fields counts as 
$\lambda^{-1/4}$, while $\hat \phi$ has the standard counting 1. 
The interaction $\frac{\lambda}{4} \phi_0^2 
\hat{\phi}^2$ therefore counts as $\sqrt{\lambda}$. In general, since 
there is always an even number of $\phi_0$ involved, 
correlation functions have an expansion in $\sqrt{\lambda}$. The rules 
are as follows: For a given correlation function expand (\ref{Zpert}) 
to the desired order in $\lambda$ using the above counting rules. 
Perform the standard Wick contractions of pairs of non-zero mode 
fields. This can be represented in terms of lines and vertices 
in the usual way. However, no Wick contractions are to be performed 
for the zero-mode fields. Instead, collect all factors of 
$\phi_0$ and compute the expectation value of $\langle \phi_0^{2n}
\rangle_0$ exactly.

As an example, we evaluate the first correction to the zero-mode 
and non-zero-mode two-point functions. For the zero-mode case, we 
have
\begin{eqnarray} 
\langle\phi_{0}\phi_{0}\rangle & = & \langle\phi_{0}^{2}\rangle_{0} 
+\langle\phi_{0}^{4}\rangle_{0} \, \times \, \  
\includegraphics[]{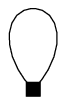}  \,
\nonumber \\[0.2cm]  & = & 
\langle\phi_{0}^{2}\rangle_{0}-\frac{\lambda}{4}\,
\langle\phi_{0}^{4}\rangle_{0}\int d^{d}x\sqrt{g}\,
\hat{G}_{{\rm free}}^{{\rm ren}}(x,x)+{\cal O}(\sqrt{\lambda}), 
\end{eqnarray}
where the black square represents the  $\frac{\lambda}{4} \phi_0^2 
\hat{\phi}^2$ vertex. If we define $\langle \phi_0 \phi_0\rangle 
\equiv 1/(V_d m_{\rm dyn}^2)$ as before in (\ref{mdyndef}) 
and denote the previously obtained leading-order expression 
(\ref{mdyn}) by  $m_{\rm dyn,0}^2$, the previous equation 
translates into 
\begin{equation} 
m_{\rm dyn}^2 = m_{\rm dyn,0}^2 
\left(1+\sqrt{\lambda} \,
\frac{3\,\Gamma(\frac{1}{4})}{2\sqrt{4!} \,\Gamma(\frac{3}{4})}
\,\sqrt{V_d}\,\hat{G}_{\rm free}^{\rm ren}(x,x)
+{\cal O}(\lambda)\right), 
\end{equation}
where $\langle \phi_0^4\rangle_0=1/(4\hat{\lambda})$ has been used. While the 
leading expression  $m_{\rm dyn,0}^2$ is unambiguous, the 
first correction depends on the UV subtraction that defines 
the coincident non-zero-mode propagator 
$\hat{G}_{\rm free}^{\rm ren}(x,x)$. To be specific, consider the case 
of four dimensions. In dimensional regularization
one needs to take $x^\prime \to x$ before expanding the $d=4-2\epsilon$  
dimensional propagator $\hat{G}_{\rm free}(x,x^\prime)$
around $d=4$, in which case
\begin{equation}
\hat{G}_{\rm free}^{\rm reg}(x,x) = \frac{H^2}{(4\pi)^2} 
\left[\frac{2}{\hat \epsilon} + 2 \ln\frac{\mu^2}{H^2} -7 + 4 \gamma_E 
\right].
\end{equation}
The $\overline{\rm MS}$-renormalized value corresponds to this expression 
with the pole term in $\frac{1}{\hat \epsilon} = \frac{1}{\epsilon}-
\gamma_E+\ln (4\pi)$ in brackets subtracted. Further, $\mu$ is the 
$\overline{\rm MS}$ renormalization scale. We note that 
the dynamical mass is not 
by itself a physical quantity. While the leading term is unambiguous, 
the first correction involving the propagation of non-zero modes is 
scheme- and scale-dependent.

The non-zero-mode two-point function is the free propagator in leading 
order. Including the ${\cal O}(\sqrt{\lambda})$ correction it 
reads
\begin{eqnarray}
\langle \hat{\phi}(x) \hat{\phi}(x^\prime) \rangle &=& 
\hat{G}_{\rm free}(x,x^\prime) 
+ \langle \phi_0^2\rangle_0
 \,  \times \, 
\parbox{85mm}{ {\centering
\includegraphics[]{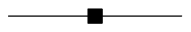}}} 
\nonumber\\[0.2cm]
&=& \hat{G}_{\rm free}(x,x^\prime) 
- \frac{\lambda}{2} \,\langle \phi_0^2\rangle_0 
\int d^dz \sqrt{g}\, \hat{G}_{\rm free}(x,z) \hat{G}_{\rm free}(z,x^\prime)
+ {\cal O}(\lambda)
\nonumber\\[0.0cm]
&=&\sum_{\vec{L}\not = \;0}H^d\,\frac{Y_{\vec L}(x) Y^*_{\vec{L}}(x^\prime)}
{H^2 L(L+d-1)} 
\left(1 - \frac{\lambda \langle \phi_0^2\rangle_0 }
{2 H^2 L (L+d-1)} + {\cal O}(\lambda)
\right).
\end{eqnarray}
The expression for the leading correction 
differs from~\cite{Rajaraman:2010xd}. The diagram computed there 
is part of the sub-leading ${\cal O}(\lambda)$ correction. In four 
dimensions the leading term equals (\ref{nonzeroprop}), and the 
leading $\sqrt{\lambda}$ correction can also be summed to give
\begin{equation}
-\frac{\lambda \langle \phi_0^2\rangle_0 }{48 \pi^2}
\left[\mbox{Li}_2(u)+\left(\frac{1}{2 u}-2\right) 
\ln(1-u) - \frac{\pi^2}{6}+\frac{1}{9}\right],
\end{equation}
where $u=1-y/4$.

\subsection*{4 Schwinger-Dyson equation}

We now return to the approaches pursued 
in~\cite{Riotto:2008mv,Garbrecht:2011gu,Serreau:2011fu} which 
are based on evaluations of the scalar-field self-energy and 
the Schwinger-Dyson equation
\begin{equation}
\Box_x G(x,x^\prime) = \frac{1}{\sqrt{g(x)}}\,\delta^{(d)}(x-x^\prime) 
+\int d^dw \,\sqrt{g(w)}\,\Pi(x,w) G(w,x^\prime).
\end{equation}
We project this equation on the zero-mode component by integrating 
over $x^\prime$ and using the identity 
\begin{equation}
\int d^d x\,\sqrt{g(x)}\,Y_{\vec{L}}(x) = \frac{\delta_{\vec{L}0}}{H^d Y_0},
\end{equation}
which follows from (\ref{orthogonality}). This results in 
\begin{equation}
0 = 1+ V_d^2 \Pi_0 G_0.
\label{SD0}
\end{equation}
In the spirit of the 2PI formalism (see below) we regard the self-energy 
as a functional of the exact propagator and derive it from the 
functional derivative 
\begin{equation}
\Pi(x,x^\prime) = 
2\,\frac{\delta \Gamma_{\rm 2PI}^{\rm rest}}{\delta G(x,x^\prime)}
\label{Pidef}
\end{equation}
of the 2PI effective action with the classical and one-loop 
term subtracted (as denoted by ``rest'')  \cite{Berges:2004yj}. 
The loop expansion of the effective action is given by 
\begin{eqnarray}
\Gamma_{\rm 2PI}^{\rm rest} &=& 
\parbox{18mm}{ {\centering \includegraphics[]{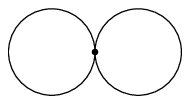}}}
 \,+\, 
\parbox{11mm}{ {\centering \includegraphics[]{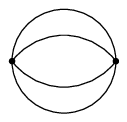}}
}
\,+\,
\parbox{11mm}{ {\centering \includegraphics[]{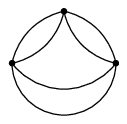}}}
\,\,+\,
\parbox{11mm}{ {\centering \includegraphics[]{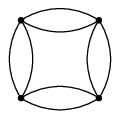}}
}
\,\,+\,
\parbox{11mm}{ {\centering \includegraphics[]{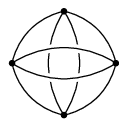}}
}  
 \,\,+\,{\cal O}(\lambda^5)  \nonumber \\[0.01cm] &  & 
{\small \hspace{7.7mm} \hbox{$\frac{1}{8}$} \hspace{18.7mm} 
\hbox{$\frac{1}{48}$} \hspace{13.8mm} \hbox{$\frac{1}{48}$}  
\hspace{13.5mm} \hbox{$\frac{1}{128}$}  \hspace{13.5mm} \hbox{$\frac{1}{32}$}}
 \quad 
\nonumber\\[0.2cm]
&=& \sum_{n=1}^{\infty} \alpha_n (-\lambda)^n 
\int d^dx_1\sqrt{g(x_1)} \ldots\int d^dx_n\sqrt{g(x_n)}
\, \underbrace{G(.,.) \ldots  G(.,.)}_{2n \,\,\rm factors}\,,
\label{2PIdiags}
\end{eqnarray}
where $\alpha_n$ denotes the combinatorial factor associated with 
a diagram, as given in the first line, and where we have given the 
explicit diagrammatic representation up to the five-loop order. 
Instead of appealing to the 
2PI formalism we could have written down the self-energy diagrams 
directly with the proviso that all internal lines are exact rather 
than free propagators.

The power-counting rules of the previous section tell us that 
the leading contribution to $\Gamma_{\rm 2PI}^{\rm rest}$ 
and $\Pi_0$ is obtained from pure zero-mode diagrams, that is, 
every full propagator is replaced by $G_0$. Since every loop 
brings one factor of $\lambda$ from the new vertex and adds 
two propagators, which each count as $1/\sqrt{\lambda}$, we 
conclude that every order in the loop expansion contributes to 
the leading term. {\em This shows that the loop expansion must 
be summed to all orders to obtain the correct 
value of the 
``dynamical mass'' in Euclidean de Sitter space. }

To see this explicitly, note that upon plugging the derivative of 
(\ref{2PIdiags}) into (\ref{SD0}), the latter equation can 
be solved for $G_0$, which yields the value of 
$m_{\rm dyn}^2$.
(More precisely, $m_{\rm dyn,0}^2$ 
since  by replacing $G$ by $G_0$ in (\ref{2PIdiags}) 
we pick up the leading term only.)
The functional derivative in (\ref{Pidef}) eliminates two 
integrations in (\ref{2PIdiags}) such that an $n$-loop self-energy 
diagram contributes $\alpha_n (-\lambda)^n V_d^{n-2} G_0^{2 n-1}$ 
to (\ref{SD0}).\footnote{Alternatively, we can substitute $G\to G_0$ in 
the expression (\ref{2PIdiags}) for $\Gamma_{\rm 2PI}^{\rm rest}$ and 
define the zero-mode self-energy as the ordinary derivative 
$2 d\Gamma_{\rm 2PI}^{\rm rest}/dG_0$. With this convention 
$\Pi_0$ as defined above must be replaced $V_d^2\Pi_0$ and the 
Schwinger-Dyson equation (\ref{SD0}) takes the form $0=1+\Pi_0 G_0$. 
This convention is adopted in Section~6 below.} 
Therefore, 
with the ansatz $G_0 = (\lambda V_d)^{-1/2} \times z$, where $z$ is a 
number to be determined,
and the explicit expression (\ref{2PIdiags}) for the loop expansion 
up to the four-loop self-energy, 
the Schwinger-Dyson equation (\ref{SD0}) turns into 
\begin{equation} 
0 = 1-\frac{z^2}{2}+\frac{z^4}{6}-\frac{z^6}{4}+\frac{5z^8}{8} 
+\ldots \equiv f(z),
\label{fz}
\end{equation}
where the term $z^{2 n}$ corresponds to the sum of $n$-loop 
self-energy diagrams. In terms of $z$, the ``dynamical mass'' 
is given by
\begin{equation}
m_{\rm dyn,0}^2 = \frac{1}{z}\times\frac{\sqrt{\lambda}}
{\sqrt{V_d}} \,\stackrel{d=4}{=}\, 
\frac{1}{z}\times\frac{\sqrt{6\lambda} H^{2}}
{4\pi}.
\label{zansatz}
\end{equation}
Keeping only the one-loop tadpole diagram in (\ref{fz}), which corresponds 
to truncation after the quadratic term, we 
obtain $z=\sqrt{2}$ and $m_{\rm dyn}^2 = \sqrt{3\lambda} H^2/(4 \pi)$ 
(in $d=4$), which coincides with the mean-field 
result \cite{Starobinsky:1994bd} and the 
one-loop result in Lorentzian de Sitter 
space \cite{Riotto:2008mv,Garbrecht:2011gu,Serreau:2011fu}. 
In the two-loop approximation, the quadratic equation for $z^2$ 
does not yield a real solution for $z$. {\em Thus, at two loops 
there is a difference between the solution of the Schwinger-Dyson 
equation in Euclidean de Sitter space and the solution to 
the corresponding equations for the closed-time-path propagators in 
Lorentzian de Sitter space} \cite{Garbrecht:2011gu}. 
Continuing to higher orders in (\ref{fz}), at three loops, we find 
$z = 1.166\ldots$, while at four loops (which includes the last term 
shown explicitly in (\ref{fz})) there is again no solution.
Thus, the loop expansion does not seem to converge to the exact 
value (\ref{mdyn}), which corresponds to
\begin{equation}
z = \sqrt{4!}\,\frac{\Gamma\left(\frac{3}{4}\right)}
{\Gamma\left(\frac{1}{4}\right)} = 1.65580\ldots.
\label{exactz}
\end{equation} 
The question arises how the exact result that is obtained easily 
from the functional integral is recovered diagrammatically. 
Clearly, we need the expansion of $f(z)$ to all orders. 
But this cannot be obtained from (\ref{2PIdiags}). While the 
integrations are trivial in the zero-mode approximation, the 
diagram topologies and computation of combinatorial factors 
become too complicated.

\subsection*{5 Zero-mode dynamics in the 2PI formalism}

In the following we exploit the 2PI 
formalism \cite{Berges:2004yj,Cornwall:1974vz}
to derive an expression that generates the perturbative 
expansion of the zero-mode self-energy to any desired order. We 
focus on the zero-mode dynamics which alone is responsible for the 
leading contributions as discussed above, and hence set 
$\hat\phi$ to zero. In this section we drop the subscript ``0'', 
since all quantities are understood to refer to the zero mode. 

The generating ``functional'' in the 2PI formalism is 
\begin{eqnarray}
Z[J,R] &=& e^{-W[J,R]} = N \int {\cal D}[\phi]\,
\exp\left(-S[\phi] - \int_x J(x) \phi(x) 
- \frac{1}{2}\int_{x,y} R(x,y) \phi(x)\phi(y)\right)
\nonumber\\[0.2cm]
&=& N \int_{-\infty}^\infty d\phi\,
\exp\left(- \tilde \lambda \phi^4  - \tilde{J} \phi 
- \frac{1}{2} \tilde{R} \phi^2 \right),
\label{Z2PI}
\end{eqnarray}
where $\tilde \lambda$ and $\tilde J$ (equal to $\tilde{J}_0$) have been 
defined in (\ref{deftilde}), $\int_x \,[...] = \int d^d x \sqrt{g(x)} \,[...]$, 
and 
\begin{equation} 
\tilde R = V_d^2 R.
\end{equation}
The simple one-dimensional integral in the second line of (\ref{Z2PI}) 
applies since the zero-mode field is constant. The exact propagator in the 
presence of the external sources can be found from the relation
\begin{equation}
\frac{dW}{d\tilde R} = \frac{1}{2}\left(\phi_{\rm cl}^2+G\right),
\label{WtoG}
\end{equation}
where $\phi_{\rm cl} = dW/d\tilde J$ is the field expectation value. 
Since the functional integral is an ordinary one-dimensional integral, 
the functional derivatives are actually ordinary derivatives. It 
follows from (\ref{Z0integrated}) that the field expectation value 
vanishes in the absence of the source $J$, that is, the symmetry 
$\phi\to-\phi$ is not spontaneously broken. This remains true 
for $J=0$ and $R=\!\!\!\!\!\!\slash\;0$. Since eventually 
we are interested in the theory in the absence of external sources 
we now put $J=0$ and consequently $\phi_{\rm cl}=0$. In this case 
we find the closed expression
\begin{equation}
Z[0,R] = e^{-W[0,R]}  = 
\frac{\sqrt{\tilde R}}{4\sqrt{2}\,\Gamma(\frac{5}{4})\,\tilde{\lambda}^{1/4}}\,
\exp\!\left(\frac{\tilde{R}^2}{32\tilde{\lambda}}\right)\,
K_{\frac{1}{4}}\!\left(\frac{\tilde{R}^2}{32\tilde{\lambda}}\right),
\end{equation}
where $K_\nu(x)$ denotes the modified Bessel function of the second kind. 
From (\ref{WtoG}) we obtain
\begin{equation}
G(R)  = 
\frac{\tilde R}{8\tilde{\lambda}}\,\left(
\frac{K_{\frac{3}{4}}\!\left(\frac{\tilde{R}^2}{32\tilde{\lambda}}\right)}
{K_{\frac{1}{4}}\!\left(\frac{\tilde{R}^2}{32\tilde{\lambda}}\right)} 
-1 \right) = \frac{H^d Y_0^2}{m^2_{\rm dyn}(R)}.
\label{mdynofR}
\end{equation}

\begin{figure}[t]
\noindent
\parbox[c]{\textwidth}{\hspace*{2cm}
\includegraphics[width=.65\textwidth]{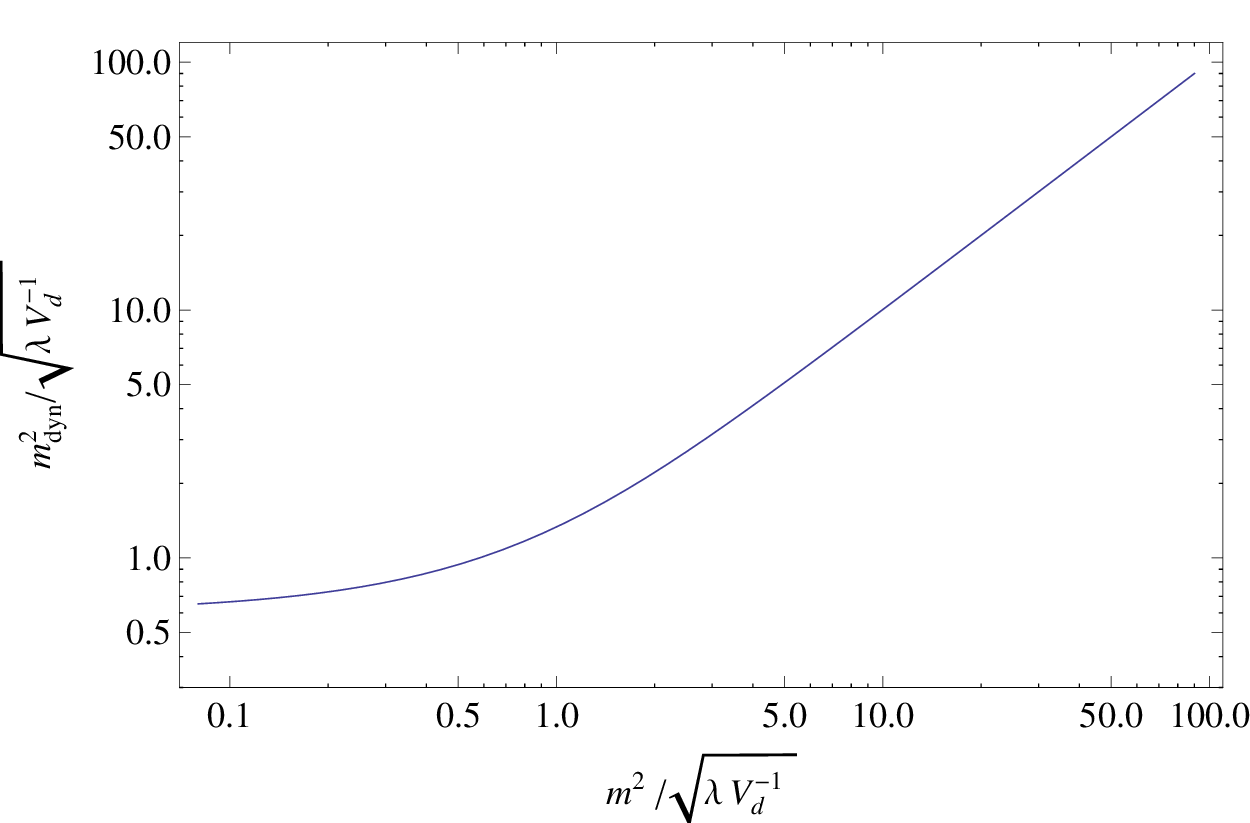}\par }
\vskip0.1cm
\parbox[t]{\textwidth}{
\caption{Dynamical mass of the zero mode as function of the 
Lagrangian mass.
\label{fig:dynamicalmass}}}
\end{figure}

The previous equation 
gives the exact zero-mode propagator or, equivalently, the 
``dynamical mass'' of the zero mode in the presence of the propagator 
source $R$. We note that since $R$ is constant, $V_d R$ has an equivalent 
interpretation as a Lagrangian mass $m^2$ for the scalar field. Hence 
(\ref{mdynofR}) provides the ``dynamical mass'' of the scalar field 
for arbitrary $m$, generalizing the expression (\ref{mdyn}) for the 
massless case \cite{Rajaraman:2010xd}. The dependence of 
$m^2_{\rm dyn}$ on $m^2$ is sketched in Figure~\ref{fig:dynamicalmass}. 
At large $m^2$ the ``dynamical mass'' asymptotes to $m^2$ with 
ordinary perturbative corrections of order $\lambda$, as should be 
expected, since the infrared enhancement that renders the zero-mode 
dynamics non-perturbative is cut off for a sufficiently massive 
scalar field. For $m^2\to 0$, the ``dynamical mass'' tends to the 
value (\ref{mdyn}). The pre-asymptotic corrections can be determined 
easily by expanding (\ref{mdynofR}) around the corresponding limits.

\subsection*{6 Diagrammatic zero-mode 
self-energy to all orders}
\label{sec:diagrammatic}

We now determine the self-energy that is needed to solve the 
Schwinger-Dyson equation. In the 2PI formalism, the 
Schwinger-Dyson equation reads  
\begin{equation}
G^{-1}(R) = G^{-1}_{\rm free} + \tilde R - \Pi(G).
\end{equation}
The inverse free propagator is $G^{-1}_{\rm free} = -12\tilde\lambda 
\phi_{\rm cl}^2$, which vanishes for $J=0$. This reflects once more 
the fact that the free propagator of the massless scalar field is 
ill-defined in the absence of external sources. Thus, 
\begin{equation}
\Pi(G) = -\frac{1}{G} + \tilde{R}(G).
\label{PiofG}
\end{equation}
It follows that the diagrammatic expansion of the self-energy is 
obtained by inverting $G(R)$ given in (\ref{mdynofR}), and 
expanding it in powers of $\lambda$.

While a closed expression for the inverse of $G(R)$ may not 
exist, we can solve for the expansion in $\lambda$ by making 
the ansatz
\begin{equation}
\tilde{R}(G) = \frac{1}{G} + a_1 \lambda V_d G + a_2  
{\lambda}^2 V_d^2 G^3 + \ldots,
\label{Ransatz}
\end{equation}
where the first term is required by (\ref{PiofG}) to obtain a 
regular perturbative expansion of $\Pi(G)$. The term 
$a_n \lambda^n V_d^n G^{2 n+1}$ represents the sum of the 
$n$-loop diagrams to the zero-mode self-energy, expressed in 
terms of the exact zero-mode propagator.
The definition of the 
expansion coefficients $a_n$ is chosen such that with the 
definition (\ref{zansatz}) of $z$ the function $f(z)$ defined 
in (\ref{fz}) is given by 
\begin{equation}
f(z) = 1-\sum_{n=1}^\infty a_n z^{2n}.
\label{expf}
\end{equation}
Plugging the ansatz (\ref{Ransatz}) into (\ref{mdynofR}) and 
matching coefficients in the expansion in $\tilde \lambda$ 
(equivalently, in $G$), we 
find the $a_n$. The first ten terms are shown in Table~\ref{tab:an}.
We note that the first four agree with 
(\ref{fz}) obtained from the combinatorial factors of 
the lowest-order Feynman-diagram topologies. We determined the 
exact coefficients up to $n=200$, which turn out to be 
rational numbers of increasing length.

\begin{table}[t]
\begin{center}%
\begin{tabular}{|c|c|c|c|c|c|c|c|c|c|c|}
\hline 
$n$ & 1 & 2 & 3 & 4 & 5 & 6 & 7 & 8 & 9 & 10\tabularnewline
\hline 
\hline 
$a_{n}$ & 
$-\frac{1}{2}$ & 
$\frac{1}{6}$ & 
$-\frac{1}{4}$ & 
$\frac{5}{8}$ & 
$-\frac{101}{48}\vphantom{\frac{\sum_{i=0}^{\infty}}{\sum_{i=0}^{\infty}}}$ & 
$\frac{279}{32}$ & 
$-\frac{8143}{192}$ & 
$\frac{271217}{1152}$ & 
$-\frac{374755}{256}$ & 
$\frac{51151939}{512}$\tabularnewline
\hline 
\end{tabular}\end{center}
\caption{The first ten series  coefficients $a_{n}$.}
\label{tab:an}
\end{table}

Inspection of the $a_n$ coefficients shows that they form a sign-alternating, 
factorially divergent series with
\begin{equation}
a_n \; \stackrel{n\to \infty}{\sim}\; 0.201... \times 
\left(-\frac{2}{3}\right)^n\,n!\times\left(1-\frac{1.7...}{n} + \ldots
\right).
\end{equation}
The divergent behaviour arises because the expansion in small $G$ 
corresponds to an expansion of $G(R)$ around $R=\infty$, 
see (\ref{Ransatz}), while the value of $m^2_{\rm dyn}$ is 
related to $G(R)$ at $R=0$. The divergent series is also the reason 
why we did not obtain a reasonable approximation to  $m^2_{\rm dyn}$ 
from the low-order approximations to the self-energy.

It remains to show that the Schwinger-Dyson approach is consistent 
with the exact result (\ref{mdyn}) for the ``dynamical mass'', which 
requires summing the divergent series. To this end we construct the 
Borel transform of $f(z)$
\begin{equation}
B[f](u) = 1-\sum_{n=1}^\infty a_n\,\frac{u^n}{n!},
\label{borelf}
\end{equation}
such that the Borel sum $F(z)$ of $f(z)$ is given by 
\begin{equation}
F(z) = \frac{1}{z^2}\int_0^\infty du\,e^{-u/z^2}\,B[f](u).
\label{borelintegral}
\end{equation}
Since the series is sign-alternating, we expect $B[f](u)$ to exhibit 
a singularity on the negative axis, but without a closed expression 
we do not know the precise singularity structure of the Borel 
transform of $f$. Given (\ref{mdynofR}), it is reasonable 
to assume that it is analytic in a vicinity of the positive real 
axis such that the Borel integral is well-defined, and to assume 
that the Borel sum $F(z)$ equals the original function $f(z)$.

Term-by-term integration of the series expansion of $B[f](u)$ 
simply returns the divergent series expansion of $f$. 
We therefore resort to a standard trick \cite{ZinnJustin:2002ru}
and construct a Pad\'{e} approximation 
from the truncated series expansion. More precisely, we use the 
first $2n$ coefficients of the expansion of $B[f]$, not counting the 
``1'' in (\ref{borelf}), and construct 
the diagonal $(n,n)$  Pad\'{e} approximant. We use this 
approximation to $B[f](u)$ in the 
Borel integral (\ref{borelintegral}) and obtain $F(z)$ by numerical 
integration. We then 
solve the equation $F(z_*)=0$, see (\ref{fz}), to determine the 
``dynamical mass''. Alternatively, we can determine the solution 
of $f(z_*)=0$ from  Pad\'{e} approximants to the expansion 
(\ref{expf}) of $f(z)$ 
directly, without going through the Borel transform. The results 
are shown in Table~\ref{tabz}. 
The solutions are seen to quickly approach the exact result 
(\ref{exactz}), especially when the Pad\'{e} approximation is 
applied to the Borel transform, in which case one-permille 
accuracy is reached already for $n=3$.  
{\em This demonstrates that the diagrammatic 
approach via the 2PI Schwinger-Dyson equation reproduces 
the path-integral result, as it must be, but only after 
summation of a divergent series expansion to all orders.}

\begin{table}[t]
\begin{center}%
\begin{tabular}{|c|c|c|c|c|c|c|c|}
\hline 
$n$ & 3 & 6 & 9 & 12 & 24 & 48 & 96 \tabularnewline
\hline 
\hline
&&&&&&& \\[-0.3cm] 
$z_*$ from $F$ & 
1.65709 & 
1.65635 & 
1.65581 & 
1.65580 & 
1.65580 & 
1.65580 & 
1.65580 \\[0.2cm]
$z_*$ from $f$ & 
1.71012 & 
1.66262 & 
1.65723 & 
1.65618 & 
1.65581 & 
1.65580 & 
1.65580 \\[0.2cm]
\hline 
\end{tabular}
\end{center}
\caption{Solution of $F(z_*)=0$ (second line) and $f(z_*)=0$ (third line) 
employing a diagonal $(n,n)$ Pad\'{e} approximation constructed from 
the first $2 n$ series coefficients $a_k$. The exact value is 
$z_*=\sqrt{4!}\,
\Gamma\left(\frac{3}{4}\right)/\Gamma\left(\frac{1}{4}\right) = 
1.65580\ldots.$}
\label{tabz}
\end{table}

\subsection*{7 \boldmath 
Stochastic approach in $d$ dimensions}

The methods applied above do not extend to Lorentzian de Sitter 
space, which is non-compact, and does not allow to identify the 
(leading) infrared dynamics with the one of a single zero-mode 
degree of freedom. However, quite some time ago Starobinsky and 
Yokoyama \cite{Starobinsky:1994bd} suggested that the long-wavelength 
part of the scalar field can be treated as a classical stochastic 
variable, which satisfies a Langevin equation with a random force 
provided by the short-wavelength modes. Here we show that this 
leads to the same value for the two-point function of the 
long-wavelength field as the zero-mode two-point function 
in Euclidean de Sitter space, in any number of dimensions $d$. This 
intriguing coincidence seems not to have been noted before.

Following  \cite{Starobinsky:1994bd} we divide the scalar field 
into $\phi=\phi_0+\hat\phi$, where $\phi_0$ contains all 
long-wavelength modes with wave number $k < \epsilon a H$ 
with $a$ the scale factor and $\epsilon \ll 1$ the parameter 
that separates long from short wave-lengths. From the field 
equation it follows that $\phi_0$ satisfies the Langevin equation 
\begin{equation}
\dot{\phi}_0(t,\vec{x}) = -\frac{1}{(d-1) H} \,V^\prime(\phi_0) 
+ f(t,\vec{x}).
\label{langevin}
\end{equation}
Here $V(\phi)$ is the scalar field potential and $f$ the 
stochastic force 
\begin{equation}
f = \dot{\hat\phi} = \epsilon a H^2 \int
\frac{d^{d-1} k}{(2\pi)^{d-1}}\,\delta(k-\epsilon a H)\, 
\left(a_k \phi_k(t) e^{i \vec{k}\cdot\vec{x}} + \mbox{c.c.}
\right)
\end{equation}
generated by the short-distance modes. At leading order, we 
can neglect the self-interaction of the short-distance modes. 
The fluctuations satisfy $\langle f(t_1,\vec{x}) f(t_2,\vec{x}) 
\rangle =\alpha \delta(t_1-t_2)$ with 
\begin{equation}
\alpha = \frac{2 \pi^{\frac{d-1}{2}}}{(2\pi)^{d-1} 
\Gamma(\frac{d-1}{2})} 
\times \frac{2^{d-3} \,[\Gamma(\frac{d-1}{2})]^2}{\pi}\times H^{d-1}
\end{equation}
The first factor arises from the volume of the $d-2$ dimensional 
momentum shell $k=\epsilon a H$, the second from the 
long-wavelength limit (since $k =  \epsilon a H \ll a H$) of the 
Bunch-Davies mode functions $\phi_k(t)$. The Fokker-Planck equation for the 
one-particle probability density $P[\varphi]$ associated with 
(\ref{langevin}) is
\begin{equation}
\frac{\partial P}{\partial t} = \frac{1}{(d-1) H} \,
\frac{\partial}{\partial\varphi}\,(V^\prime(\varphi) P) 
+\frac{\alpha}{2}\,\frac{\partial^2}{\partial\varphi^2}\,P,
\label{fokkerplanck}
\end{equation}
which admits the stationary late-time solution 
\begin{equation}
P[\varphi] = N \exp\left(-\frac{2}{(d-1)\alpha H}\,V(\varphi)\right),
\end{equation}
in terms of which the two-point function of the (constant) long-wavelength 
field is given by
\begin{equation}
\langle \phi_0\phi_0 \rangle = N\int_{-\infty}^\infty d\varphi \,
\varphi^2  \exp\left(-\frac{2}{(d-1)\alpha H}\,V(\varphi)\right).
\end{equation}
This precisely agrees with (\ref{phisquared}) (for $V(\phi) = 
\frac{\lambda}{4!}\phi^4$) provided the dissipation and fluctuation 
coefficients in the Fokker-Planck equation are related to the volume of 
$d$-dimensional Euclidean de Sitter space with radius $1/H$ by 
\begin{equation}
(d-1) H \times\frac{\alpha}{2} = \frac{1}{V_d},
\end{equation}
which can be easily verified. {\em Hence, the long-wavelength 
two-point functions (and therefore ``dynamical masses'') are the same, as 
claimed.} One may wonder why the result could be derived from zero-mode 
dynamics alone in Euclidean de Sitter space, while the fluctuations 
originated from the short-wavelength modes in the stochastic 
approach. However, the result from the latter is independent of 
$\epsilon$ in the above approximation, as it should be, and the 
stochastic force is generated by wave numbers $k=\epsilon a H$. 
We can take $\epsilon$ arbitrarily small and conclude that in the 
leading approximation the main contribution to the stochastic 
force can be assumed to originate from the boundary between long- 
and short wavelengths, which can be taken to be deep in the infrared.

We note that the stochastic approach can be derived rigorously 
from the full quantum dynamics in the leading logarithmic 
approximation in $a$ \cite{Prokopec:2007ak}, which is equivalent 
to keeping the leading infrared-enhanced terms in Euclidean 
de Sitter space. But unlike the Euclidean case discussed in 
the present paper, a systematic method for calculating corrections 
around the Lorentzian result of  \cite{Starobinsky:1994bd} 
is not known.

\subsection*{8 Conclusion}

In this paper we considered the perturbative treatment of the minimally 
coupled, massless, self-interacting scalar field in Euclidean de 
Sitter space. Generalizing the work of Rajaraman \cite{Rajaraman:2010xd}, 
we obtained the dynamical mass $m^2\propto \sqrt{\lambda} H^2$ of 
the scalar for non-vanishing Lagrangian masses, as well as the first 
perturbative quantum correction in the massless case, and developed the 
rules of a systematic perturbative expansion, which after treating the 
zero-mode non-perturbatively, goes in powers of $\sqrt{\lambda}$. 
We then showed how the dynamical mass can be recovered from the 
summation of the diagrammatic expansion of the self-energy and a 
consistent solution of the Schwinger-Dyson equation. This clarifies 
the relation between the path-integral and diagrammatic treatment, 
and implies that solutions based on truncations of the loop expansion 
can at best be approximate. With the proper exact treatment of the 
zero mode, the reorganized perturbative expansion is free from 
infrared divergences, which are present for the free, minimally 
coupled scalar field in Euclidean de Sitter space. The interacting, 
massless field is therefore well-defined, and the rules for generating 
the systematic perturbative expansion are almost as simple as 
the standard rules for the massive case.

What this implies for Lorentzian de Sitter space is much less clear. 
We showed that the long-wavelength mode two-point function computed 
in the stochastic approach of \cite{Starobinsky:1994bd} coincides 
with the exact Euclidean result in leading order in the expansion in  
$\sqrt{\lambda}$. This strongly suggests to us that the dynamical 
mass of the self-interacting scalar field in de Sitter space 
can be obtained by some sort of analytic continuation from the 
Euclidean, up to higher-order corrections. It would be very 
interesting to derive this result diagrammatically, in the 
spirit of \cite{Garbrecht:2011gu}, and to understand how to 
develop a systematic expansion in  Lorentzian de Sitter space. 

\vspace*{0.5em}
\noindent
\subsubsection*{Acknowledgement}
We thank B.~Garbrecht, T.~Prokopec and G.~Rigopoulos for discussions.
This work is supported in part by the Gottfried Wilhelm 
Leibniz programme of the Deutsche Forschungsgemeinschaft (DFG).

\bibliographystyle{elsarticle-num}

\end{document}